\documentclass[pdflatex,sn-mathphys-num]{sn-jnl}

\usepackage{graphicx}%
\usepackage{multirow}%
\usepackage{amsmath,amssymb,amsfonts}%
\usepackage[ruled,vlined,linesnumbered]{algorithm2e}            
\usepackage{amsthm}%
\usepackage{mathrsfs}%
\usepackage[title]{appendix}%
\usepackage{xcolor}%
\usepackage{textcomp}%
\usepackage{manyfoot}%
\usepackage{booktabs}%
\usepackage{listings}%
\usepackage{booktabs}
\usepackage{makecell}
\usepackage{multicol}
\usepackage{multirow}
\usepackage{tabularx}
\usepackage{caption}
\usepackage{float}
\usepackage{afterpage}
\usepackage{longtable}
\usepackage{xltabular}
\usepackage{enumitem}
\usepackage{tabularx}
\usepackage{geometry}
\usepackage{comment}
\usepackage{threeparttable}
\usepackage{array} 
\usepackage{subcaption}
\usepackage{tikz}
\usetikzlibrary{trees, positioning}
\usepackage{adjustbox}
\usepackage{rotating}

\newcolumntype{C}[1]{>{\centering\arraybackslash}p{#1}}
\usepackage{booktabs} 
\newcolumntype{R}[1]{>{\raggedright\arraybackslash}p{#1}}
\usepackage{caption} 

\theoremstyle{thmstyleone}%
\theoremstyle{thmstyletwo}%

\theoremstyle{thmstylethree}%

\begin{document}

\title[Article Title]{Comparative algorithm performance evaluation and prediction for the maximum clique problem using instance space analysis}


\author*[1]{\fnm{Bharat} \sur{S. Sharman}}\email{sharmanb@mcmaster.ca}

\author[2]{\fnm{Elkafi} \sur{Hassini}}\email{hassini@mcmaster.ca}


\affil*[1]{\orgdiv{School of Computational Science and Engineering}, \orgname{McMaster University}, \orgaddress{\city{Hamilton}, \postcode{L8S 4E8}, \state{ON}, \country{Canada}}}

\affil[2]{\orgdiv{DeGroote School of Business}, \orgname{McMaster University}, \orgaddress{\city{Hamilton}, \postcode{L8S 4E8}, \state{ON}, \country{Canada}}}


\abstract{The maximum clique problem, a well-known graph-based combinatorial optimization problem, has been addressed through various algorithmic approaches, though systematic analyses of the problem instances remain sparse. This study employs the instance space analysis (ISA) methodology to systematically analyze the instance space of this problem and assess \& predict the performance of state-of-the-art (SOTA) algorithms, including exact, heuristic, and graph neural network (GNN)-based methods. A dataset was compiled using graph instances from TWITTER, COLLAB and IMDB-BINARY benchmarks commonly used in graph machine learning research. A set of 33 generic and 2 problem-specific polynomial-time-computable graph-based features, including several spectral properties, was employed for the ISA. A composite performance measure incorporating both solution quality and algorithm runtime was utilized. The comparative analysis demonstrated that the exact algorithm Mixed Order Maximum Clique (MOMC) exhibited superior performance across approximately 74.7\% of the instance space constituted by the compiled dataset. Gurobi \& CliSAT accounted for superior performance in 13.8\% and 11\% of the instance space, respectively. The ISA-based algorithm performance prediction model run on 34 challenging test instances compiled from the BHOSLIB and DIMACS datasets yielded top-1 and top-2 best performing algorithm prediction accuracies of 88\% and 97\%, respectively.}

\keywords{Combinatorial optimization, Instance space analysis, Maximum clique problem, Graph neural networks, Algorithm performance comparison}



\maketitle

\section{Introduction}\label{sec:intro}

With the advent of the big data era, the volume, velocity, and variety of data in almost all fields of science has increased significantly (\cite{hansen2014big}). The field of combinatorial optimization has also seen a growth in the size and complexity of problems (\cite{weinand2022research}). Combinatorial optimization problems (COPs) involve selecting a subset of objects from a finite set such that the selected subset optimizes an objective function (\cite{du2022introduction}). COPs are of immense theoretical and practical importance. For example, it is well known that several COPs, such as the traveling salesperson problem (TSP) are $\cal{NP}$-complete and are widely used to formulate and solve problems in several areas such as supply chain management, logistics, bioinformatics, and semiconductor chip manufacturing, to name but a few. \\
Many COPs, including graph coloring, maximum clique, maximum independent set, minimum vertex cover, vehicle routing (VRP), and location routing (LRP), can be formulated using graph representations. Several algorithms have been proposed to solve such graph-based COPs. However, there is no single algorithm that can solve every case of a problem optimally according to the Theorem of No-Free Lunch of \cite{wolpert1997no}. Therefore, the algorithm that is best suited to solve a problem instance needs to be determined based on a systematic evaluation of the structure of the problem instance. This becomes especially important for graph-based COPs as many of them are NP-complete, and therefore the time required to solve them to optimality increases exponentially with the size of the problem instance in the worst-case scenario. \\
Performance evaluation of most algorithms has traditionally been based on specific benchmark datasets. Although this approach facilitates direct comparisons by evaluating algorithms on identical problem instances, it may represent only a limited subset of the broader instance space. Consequently, such evaluations can inadvertently favor certain algorithms while disadvantaging others, depending on the alignment between sample instances and the characteristics of the evaluated algorithms. Furthermore, algorithm performance on the benchmark datasets is summarized using average metrics across all problem instances. Such an aggregation obscures variations in performance on individual problem instances, potentially concealing important insights into the relative strengths and weaknesses of different algorithms. \\
A promising approach to alleviate the aforementioned challenges in comparing algorithm performance is the Instance Space Analysis (ISA) methodology proposed by \cite{smith2023instance}. ISA involves mapping problem instances to a two-dimensional instance space, enabling the identification of underrepresented regions within the instance datasets. Additionally, ISA allows for the analysis of algorithm performance across the entire instance space, enabling the identification of algorithms that perform optimally in specific regions and on individual problem instances. \\
This study focuses on the Maximum Clique Problem (MCP), which involves finding the largest fully connected subgraph in a given graph (\cite{wu2015review}). The decision variant of the MCP was among the 21 problems originally classified as NP-complete in the seminal work of (\cite{karp1972}). Due to its theoretical significance and numerous practical applications, the MCP has garnered extensive research interest. Several algorithmic approaches have been proposed to solve the MCP such as exact algorithms with branch and bound (B\&B) approaches (\cite{li2017minimization}, \cite{li2010efficient}, \cite{jiang2016combining}, \cite{san2019new}, \cite{san2023clisat}), greedy and local search approaches (\cite{wu2012multi}, \cite{wang2016two}, \cite{wang2020sccwalk}), Monte Carlo methods (\cite{angelini2021mismatching}) and more recently, graph neural networks (GNN) (\cite{karalias2020erdos}, \cite{min2022can}, \cite{sanokowski2024variational}) and quantum computing approaches (\cite{haverly2021implementation}, \cite{li2024parameter}, \cite{pelofske2019solving}).\\ 
These algorithms have their own distinct strengths and limitations, and no one algorithm consistently outperforms all others in all instances of the problem. This underscores the importance of evaluating algorithmic performance across the instance space to determine which algorithms perform best in specific regions. Such an approach enables the selection of an appropriate algorithm for a given problem instance, thereby avoiding deploying expensive computational resources and spending time on algorithms that may not be a good fit for a problem instance based on its location in the instance space. There have been several reviews (\cite{pardalos1994maximum}, \cite{bomze1999maximum}, \cite{wu2015review}) as well as studies (\cite{feige2004approximating}, \cite{engebretsen2003towards}, \cite{zuckerman2006linear}, \cite{chalermsook2020gap}) on the complexity of the MCP. However, to the best knowledge of the authors, there have yet been no systematic studies characterizing the performance of different algorithms in a diverse and representative instance space of the MCP. The results of such a study can be useful for practitioners in the industry or researchers in selecting suitable algorithms. It can also inspire further research into the underlying reasons for the comparative advantages of certain algorithms. The main contributions of this study are therefore the following:

\begin{itemize}

    \item Instance space analysis for algorithm characterization: We analyze the performance of five MCP algorithms using the Instance Space Analysis (ISA) framework,  identifying regions in the instance space where these algorithms exhibit superior performance.
    \item Generic graph-based features: We emphasize generic graph-based features to characterize algorithm performance. Out of the 35 features that we used, 33 were generic and only 2 were specific to the MCP. This approach generalizes the methodology, making it applicable to investigate and compare other graph-based COPs.
    \item Composite hardness metric: We introduce a composite hardness metric that integrates solution quality and runtime, providing a more comprehensive evaluation of algorithm performance.
    \item Algorithm performance prediction: We predict algorithm performance on 34 challenging instances of the BHOSLIB and DIMACS dataset through the trained ISA-based support vector classifier (SVC) model and demonstrate high prediction accuracies and significant time savings achieved through our methodology.

These contributions aim to provide actionable insights for both academic and industrial practitioners while fostering further advances in the study and application of algorithms for the MCP and combinatorial optimization in general. 
\end{itemize}

\section{The maximum clique problem}
\label{sec:mcp}

\subsection{Definitions and notations}

Let $G = (V,E)$ be an undirected and unweighted graph where V is the set of nodes and E is the set of edges of the graph. The vertex cardinality of a graph is the number of vertices and is indicated by $|V|$ and the edge cardinality of a graph is the number of edges and is denoted by $|E|$. For a subset of nodes $S \subseteq V$, the set $G(S) = G(S,E \subseteq S \times S)$ is known as the sub-graph induced by S. A complete graph is a graph where every pair of vertices is adjacent, i.e. $\forall (i,j) \in V$, if $i \neq j$, then $(i,j) \in E $. A clique $C$ is a complete subgraph of $G$. A clique is called maximal if it is not a subset of a larger clique in $G$. A clique is said to be the maximum clique if its vertex cardinality is the highest among all the cliques of the graph. The MCP can be formulated as the following integer linear programming problem:

\begin{equation}
\label{eq:mcp_ilp_formulation}
\begin{aligned}
    \text{Maximize } & \sum_{i \in V} x_i \\
    \text{subject to: } & x_i + x_j \leq 1 \quad \forall (i, j) \notin E \\
    & x_i \in \{0, 1\} \quad \forall i \in V
\end{aligned}
\end{equation}

where \( x_i \) is a binary variable that indicates whether the vertex \( i \) is included in the clique (\( x_i = 1 \)) or not (\( x_i = 0 \)). The inequality constraint in Equation \ref{eq:mcp_ilp_formulation} specifies the condition that if there is no edge between two nodes in the graph, they cannot both be members of a clique. The clique number $\omega (G)$ is the number of vertices in the maximum clique of G. In addition to this classic version of the MCP defined on an unweighted graph, weighted versions of the MCP also exist. If each node $i$ of a graph has a weight $w_{i}$ associated with it, these weights can collectively be represented by a weight vector $ \vec w \in R^{N} $. The Maximum Weighted Clique Problem (MWCP) requires finding the clique with the maximum sum of node weights. In an analogous manner, the maximum weighted clique problem can also involve finding the clique with the maximum sum of edge weights instead of node weights. The focus of this work is on the classic version of the MCP.
        
\subsection{Algorithmic approaches for solving the MCP}
\label{subsec:algorithmic_approaches_mcp}
Algorithmic approaches for solving the MCP can be classified into four broad classes: exact, heuristic, machine learning-based, and quantum-based algorithms. \\
Among these, exact algorithms have the longest history of development and have been primarily based on the B\&B technique. Comprehensive reviews of exact algorithms for the MCP and the B\&B technique are provided by \cite{wu2015review} and \cite{lawler1966branch}, respectively. We now outline several notable exact algorithms introduced in recent years. \cite{li2017minimization} demonstrated that the efficacy of incremental MaxSAT reasoning was effective in reducing branching operations and proposed the Mixed Order Maximum Clique (MOMC) algorithm. By combining static and dynamic branching strategies within the B\&B framework, MOMC outperformed other algorithms relying solely on a single branching strategy when tested on several random graphs and DIMACS instances. \cite{jiang2017exact} proposed the Weighted Large Maximum Clique (WLMC), which is particularly effective for large sparse vertex-weighted graphs. The algorithm incorporates two novel steps: an initial vertex ordering to reduce the size of the graph and incremental vertex splitting to reduce the number of branches. \cite{san2019new} proposed an algorithm for MWCP based on independent set cover. They proposed a new bounding function that achieved a good trade-off between the pruning potential and the computing effort. \cite{szabo2018different} approached the MCP by representing it as a sequence of $k$-clique decision problems. This transformation provided additional pruning opportunities within the B\&B tree compared to the original maximization formulation of the MCP. Their numerical experiments supported the effectiveness of this approach. \cite{san2023clisat} presented the CliSAT algorithm, incorporating two novel bounding techniques. The first utilizes graph-coloring procedures and partial MaxSAT problems within the branching scheme, while the second applies a filtering phase based on constraint programming and domain propagation. CliSAT achieves SOTA performance on large graphs containing numerous interconnected large cliques.\\  
Heuristic algorithms play a pivotal role in advancing both the theoretical development and practical applications of COPs. Unlike exact algorithms, heuristic algorithms do not guarantee the discovery of optimal solutions. However, their approximate solutions can be valuable, particularly for problems that are computationally infeasible for exact algorithms. Heuristics use domain-specific knowledge, predefined rules, problem structure, or a combination of these elements to produce their solutions. Greedy search and local search are two of the well-researched heuristic algorithms for the MCP. Greedy search solves problems iteratively by decomposing them into a sequence of sub-problems and selecting the most favorable solution at each step. In the context of MCP, \cite{karp1976probabilistic} introduced one of the most notable greedy algorithms that can find a clique up to the size of $|C| = log_{2}N$ in a graph comprising of $N$ nodes. This algorithm iteratively selects a random node, removes approximately half of the unconnected nodes, and repeats this process until the graph is fully explored. Building on Karp's work, \cite{brockington1993camouflaging} proposed improvements to the node selection mechanism, prioritizing nodes with higher connectivity. This enhancement significantly improved performance, particularly in smaller graphs. \cite{marino2023hard} studied the structure of the MCP as a function of N, the graph size, and K, the clique size sought and found that it displayed a phase transition behavior. They introduced a hybrid method combining iterative and exhaustive search strategies, which outperformed other greedy approaches while maintaining low computational complexity. Local search methods are based on a search strategy guided by a defined search space, a neighborhood function, and an evaluation function. Among these, the FastWClq algorithm proposed by \cite{cai2016fast} has demonstrated superior performance on large graphs. FastWClq integrates clique formation with graph reduction. In each cycle, a clique starts as empty, employing pruning to halt non-optimal constructions. If a superior clique is identified, it updates the best clique while applying reduction rules to decrease graph size. \\
Machine learning algorithms, particularly graph neural networks (GNNs), have emerged as a powerful class of solution methods for tackling COPs (\cite{cappart2023combinatorial}, \cite{bengio2021machine}). GNNs are uniquely suited for graph-based COPs due to their ability to aggregate information and perform predictive analysis across various levels, including nodes, edges, sub-graphs, and graphs. In the context of the Maximum Clique Problem (MCP), \cite{karalias2020erdos} introduced the Erdos Goes Neural (EGN), an unsupervised GNN-based framework inspired by Erdos' probabilistic method. EGN guarantees solutions of certified quality while adhering to the constraints of the MCP. Building on this probabilistic approach, \cite{min2022can} developed the Hybrid Graph Scattering (HGS) framework, which also employs unsupervised learning and the probabilistic method. HGS addresses the oversmoothing issue commonly associated with GNNs by incorporating a scattering transform, enabling it to achieve superior performance to other GNN models, including EGN, while having only 0.1\% of the number of model parameters. In contrast to the probabilistic methods used in EGN and HGS, \cite{sanokowski2024variational} proposed an auto-regressive approach. This method captures statistical dependencies between solution variables, diverging from the assumption of statistical independence inherent in probabilistic methods. The auto-regressive approach demonstrated superior performance over EGN and other solvers across several MCP benchmark datasets.\\
Quantum computing (QC) approaches have increasingly been applied to solve COPs. The primary QC methodologies used in this context include the Quantum Minimum Finding Algorithm (or the circuit model), Quantum Annealing (QA), and Quantum Approximate Optimization Algorithm (QAOA). The circuit model for solving MCP based on Grover's search algorithm (\cite{grover1996fast}) has been used to find maximum cliques in a graph of size N with a time complexity of $ \sim O (|N| \sqrt(2^{|N|}))$ (\cite{bojic2012quantum}). This algorithm can be implemented on devices using open source tools such as Qiskit (\cite{haverly2021implementation}). QA is based on analog computation and is suited for solving Quadratic Unconstrained Binary Optimization (QUBO) problems heuristically. As the Maximum Independent Set Problem (MISP) is the complementary problem of the MCP and can be conveniently rendered as a QUBO, this approach has been used to solve the MCP (\cite{chapuis2017finding}). However, results have shown no quantum speedup over classical algorithms for random graphs that fit the D-Wave 2X machine. Significant speedups were observed only for problem instances specifically designed to align well with the quantum annealer’s qubit interconnection network, emphasizing the critical role of problem structure. \cite{pelofske2023solving} extended QA to midsized graphs (up to 120 nodes and 6395 edges) using a hybrid method that combined parallel quantum annealing with graph decomposition. QAOA, often considered the digital equivalent of QA, implements QA on gate-based quantum devices (preprint-\cite{farhi2014quantum-arxiv}). Using this approach, \cite{ebadi2022quantum} reported solving unit-disk graph MISPs with up to 80 vertices employing Rydberg atom arrays comprising up to 289 coupled qubits in two dimensions.\\
This review highlights the diversity of algorithmic approaches for solving the MCP. Furthermore, the algorithms utilize different datasets, making it difficult to compare and systematically characterize the algorithmic performance for the MCP. Addressing these challenges is essential for advancing our understanding of algorithm efficacy and determining their applicability to specific problem instances.

\section{Instance space analysis}
\label{sec:isa}
\begin{figure}[h]
    \centering
    \includegraphics[width=0.8\textwidth]{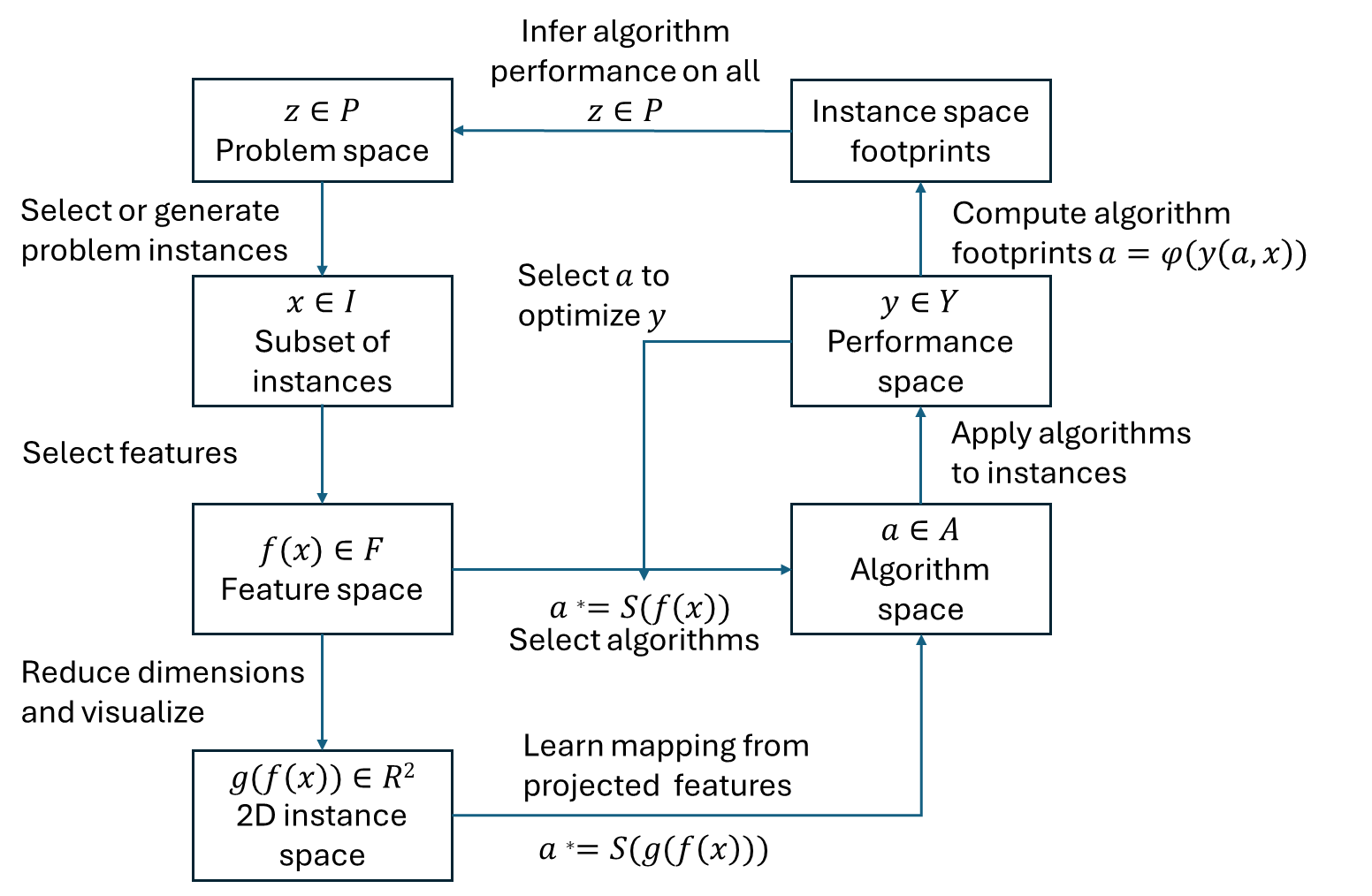}
    \caption{Instance Space Analysis methodological framework that extends Rice’s Algorithm Selection Problem}
    \label{fig:ISA_Framework}
\end{figure}

The objective of Instance Space Analysis (ISA) is to address the challenge of comparing and characterizing algorithmic performance independently of the choice of test dataset. The ISA framework introduced by \cite{smith2023instance} is based on the foundational algorithm selection framework proposed by \cite{rice1976algorithm}. The ISA methodological framework is shown in Figure \ref{fig:ISA_Framework}.     

The ISA framework begins by taking the instances of a problem as inputs and constructs a predictive mapping between the instances and the performance of the algorithm using an iterative learning-based approach. It transforms the selected features into a two-dimensional representation through dimensionality reduction techniques, allowing visualization of regions where specific algorithms exhibit superior or inferior performance relative to others. In addition, it also computes the footprint of an algorithm in the instance space, which is a quantitative measure of the general effectiveness of an algorithm for the problem. A larger footprint indicates broader applicability and effectiveness across various problem instances compared to algorithms with smaller footprints. ISA has been successfully applied to evaluate algorithm performance for a variety of problems such as classification (\cite{munoz2018instance}), regression (\cite{munoz2021instance}), job shop scheduling (\cite{strassl2022instance}), knapsack (\cite{smith2021revisiting}), multi-objective optimization (\cite{yap2022informing}), maximum flow (\cite{alipour2023enhanced}) among others. 

However, to the best of the authors' knowledge, no prior study has applied the ISA methodology to the MCP to explore its instance space and conduct a comparative evaluation of algorithm performance. This study aims to address this research gap by leveraging ISA to gain deeper insights into algorithm performance on the MCP. 

\section{Instance space analysis for the MCP}
\label{sec:isa_for_mcp}

This section outlines the methodology adopted to perform the ISA on the MCP. The selection of algorithms and problem instances, feature extraction and selection, hardness metric formulation, and key aspects of the ISA workflow specific to this analysis are now discussed. 
\subsection{Algorithms}
As highlighted in Section \ref{subsec:algorithmic_approaches_mcp}, a wide variety of algorithms have been developed to solve the MCP. For this study, five algorithms were selected, namely Gurobi's Mixed Integer Linear Programming Solver (Gurobi), CliSAT, Mixed Order Maximum Clique (MOMC), Fast Solving Weighted Clique (FastWClq) and Hybrid Geometric Scattering (HGS). Gurobi, CliSAT, and MOMC are all well-known B\&B-based exact algorithms while FastWClq and HGS represent SOTA heuristic and unsupervised GNN-based algorithmic approaches for the MCP respectively. To the best knowledge of the authors, there has been no study that has systematically compared the performance of exact, heuristic and GNN-based algorithms for the MCP. 

\subsection{Problem Instances}
\begin{figure}[ht!]
    \centering
    \includegraphics[width=1.0\textwidth]{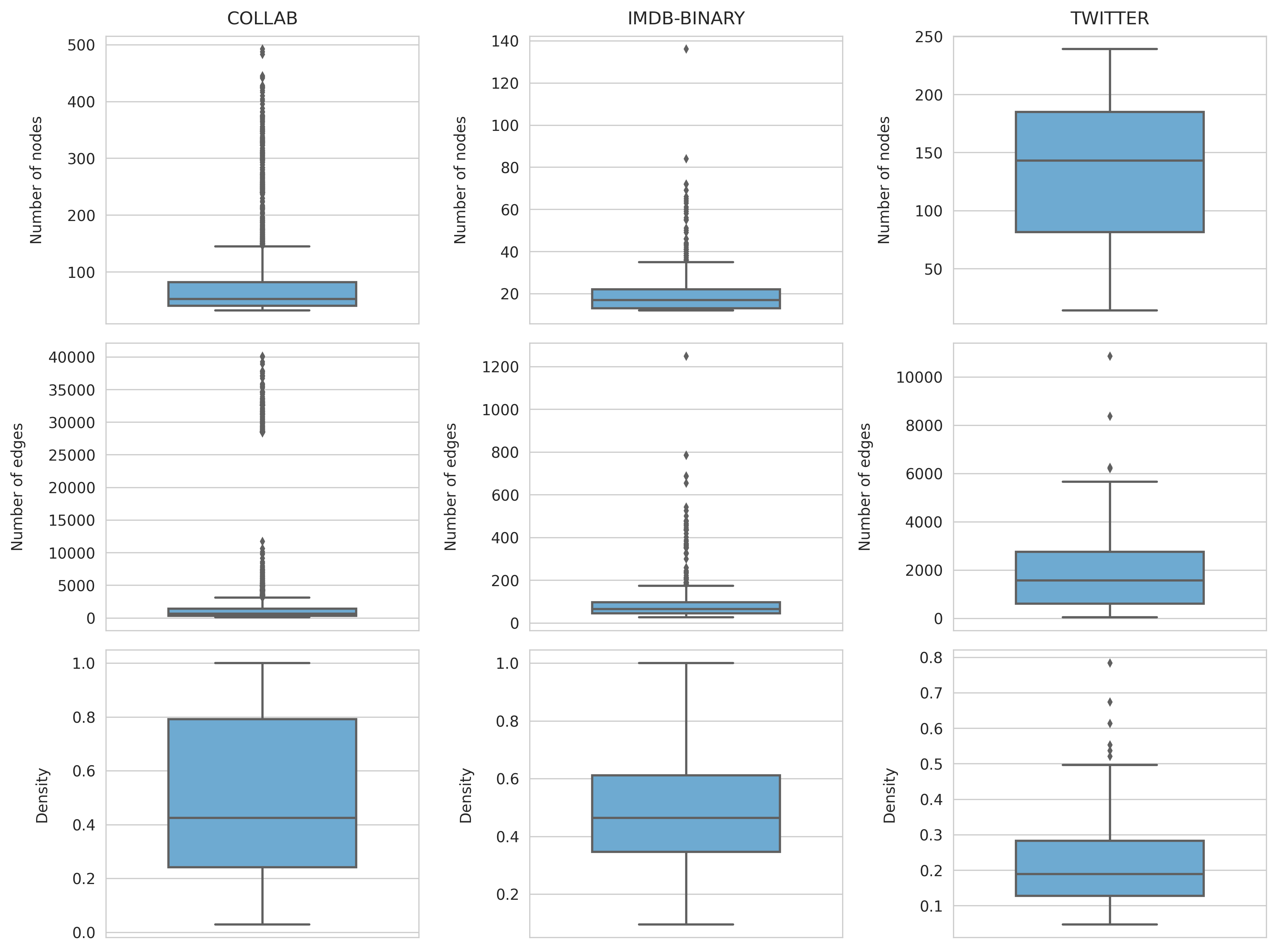}
    \caption{Node count, edge count, and density distribution of 6138 graph instances used for performing the initial instance space analysis}
    \label{fig:ISA_graphs_summary}
\end{figure}

The analysis was performed on a dataset of 6138 graph instances. A summary of these instances is provided in Table \ref{tab:summary_of_graphs_for_isa}. 1 instance each from the TWITTER, COLLAB, and the IMDB-BINARY datasets was not included, as their graph features could not be computed either because of discontinuities in the graphs or if the computation time for features exceeded two minutes per instance. Only 3 out of 6138 instances took more than 1 minute for the computation of their features. The datasets TWITTER, COLLAB, and IMDB-BINARY were considered because they are commonly used to benchmark the performance of GNN-based algorithms for the MCP. 141 out of 973 instances of the TWITTER dataset were selected as the remaining instances were used as part of the training and validation dataset for training the HGS algorithm. Even though HGS is an unsupervised algorithm, its weights were initially trained using a dataset as per the recommendation of (\cite{karalias2020erdos}) in their work.  Performing ISA based on this compilation can not only illustrate areas of the instance space where GNN-based algorithms have their comparative advantages and disadvantages, but also show how they compare against exact B\&B-based and heuristic algorithms on these graph instances. Figure \ref{fig:ISA_graphs_summary} illustrates the distributions of node count, edge count, and density for the graph instances in the three datasets.

\vspace{2\baselineskip} 
\begin{table}[h]
    \centering
    \caption{Summary of graph instances datasets used for performing instance space analysis}
    \begin{tabularx}{\textwidth}{>{\centering\arraybackslash}X >{\centering\arraybackslash}X >{\centering\arraybackslash}X}
        \hline
        \textbf{Dataset Name} & \textbf{Source} & \textbf{Number of Graphs} \\
        \hline
        TWITTER & [dataset] \cite{snapnets} & 140 \\
        \hline
        COLLAB & [dataset] \cite{morris2020tudataset} & 4999 \\
        \hline
       IMDB-BINARY & [dataset] \cite{morris2020tudataset} & 999 \\
        \hline
    \end{tabularx}
    \label{tab:summary_of_graphs_for_isa}
\end{table}
\vspace{2\baselineskip} 

\subsection{Features}
The 35 features considered for ISA in this study and their corresponding computational complexities are listed in Table \ref{tab:graph_features_used_for_isa}. Several of these graph-based generic features have been selected based on prior works of \cite{smith2014exploring} and \cite{smith2015generating}  who analyzed the spectral
properties of graphs to evaluate algorithm performance on the graph coloring problem. Numerous spectral features have been shown to be computed in polynomial time \cite{smith2012measuring}, a finding corroborated by this study, since all features listed in Table \ref{tab:graph_features_used_for_isa} were similarly computed in polynomial time. Notably, 33 of these 35 features are generic and applicable across a variety of graph-based COPs as opposed to being tailored for specific COPs. Consequently, their use offers the potential to gain valuable insights into algorithm performance across a wide range of problem classes. Two important features specific to the maximum clique problem, namely the gap between a graph's chromatic number \& a greedy clique number, and the K-core number, were computed using the built-in heuristic algorithms of the Networkx library. 

\subsection{Algorithm performance measure}

The performance measure for ISA in this study is a composite measure integrating solution quality and runtime. It is defined as follows: (\ref{eq:hardness_criterion_isa}): 
\begin{equation}
\label{eq:hardness_criterion_isa}
y_{\min}
  \;=\;
  \frac{
        \displaystyle
        \frac{\text{time taken (s)}}%
             {\max\!\bigl(\text{time taken (s) by all algorithms for the instance}\bigr)}
       }%
       {
        \displaystyle
        \frac{\text{clique size obtained}}%
             {\max\!\bigl(\text{clique size obtained by all algorithms for the instance}\bigr)}
       }
\end{equation}

The metric is structured so that lower values signify superior algorithm performance on an instance. For example, algorithms A and B, computing cliques of sizes 80 and 100 in 80 and 160 seconds respectively, yield performance measures of $\frac{5}{8}$ and 1. Thus, A, with a lower metric value, outperforms B. This allows comparison of algorithms with varying solution quality and runtime using a unified standard.      

\afterpage{%
\clearpage
\begin{longtable}{R{12cm} R{3cm}}
\caption{Graph‑based features used for performing instance‑space analysis and their computational complexity}
\label{tab:graph_features_used_for_isa} \\[2pt]
\hline
\textbf{Feature name} & \textbf{Computational complexity}\\
\hline
\endfirsthead

\multicolumn{2}{c}{{\bfseries \tablename\ \thetable{} -- continued from previous page}}\\
\hline
\textbf{Feature name} & \textbf{Computational complexity}\\
\hline
\endhead

\hline
\multicolumn{2}{r}{\small Continued on next page}\\
\hline
\endfoot

\hline
\endlastfoot
Number of nodes & \(O(|V|)\)\\
Number of edges & \(O(|E|)\)\\
Density & \(O(1)\)\\
Girth & \(O(|V|\cdot|E|)\)\\
Diameter & \(O(|V|\cdot|E|)\)\\
Median of the betweenness centralities of all nodes & \(O(|V|\cdot|E|)\)\\
Median of the closeness centralities of all nodes & \(O(|V|\cdot|E|)\)\\
Median of the degree centralities of all nodes\(^\dagger\) & \(O(|E|+|V|\log|V|)\)\\
Median of the eigenvector centralities of all nodes\(^\ddagger\) & \(O(k\,|E|)\)\\
Standard deviation of the betweenness centralities of all nodes & \(O(|V|\cdot|E|)\)\\
Standard deviation of the closeness centralities of all nodes & \(O(|V|\cdot|E|)\)\\
Standard deviation of the degree centralities of all nodes\(^\dagger\) & \(O(|E|+|V|\log|V|)\)\\
Standard deviation of the eigenvector centralities of all nodes\(^\ddagger\) & \(O(k\,|E|)\)\\
Median of the degree of all nodes & \(O(|E|)\)\\
Standard deviation of the degree of all nodes & \(O(|E|)\)\\
Median of the median degrees of the neighbours of a node & \(O(|E|)\)\\
Standard deviation of the median degrees of the neighbours of a node & \(O(|E|)\)\\
Median of the geodesic distance of all node pairs & \(O(|V|\cdot|E|)\)\\
Standard deviation of the geodesic distance of all node pairs & \(O(|V|\cdot|E|)\)\\
Global clustering coefficient\(^\S\) & \(O(|V|+|E|)\)\\
Proportion of even closed walks to all closed walks & \(O(|V|^{3})\)\\
Spectral radius & \(O(|V|^{3})\)\\
Laplacian spectral radius & \(O(|V|^{3})\)\\
Energy & \(O(|V|^{3})\)\\
Standard deviation of all the eigenvalues of the adjacency matrix & \(O(|V|^{3})\)\\
Smallest non‑zero eigenvalue of the Laplacian matrix & \(O(|V|^{3})\)\\
Second smallest non‑zero eigenvalue of the Laplacian matrix & \(O(|V|^{3})\)\\
Second largest eigenvalue of the Laplacian matrix & \(O(|V|^{3})\)\\
Smallest eigenvalue of the adjacency matrix & \(O(|V|^{3})\)\\
Second smallest eigenvalue of the adjacency matrix & \(O(|V|^{3})\)\\
Second largest eigenvalue of the adjacency matrix & \(O(|V|^{3})\)\\
Difference between the largest and the second largest eigenvalues of the adjacency matrix & \(O(|V|^{3})\)\\
Difference between the largest and the smallest eigenvalues of the Laplacian matrix & \(O(|V|^{3})\)\\
K‑core number & \(O(|V|+|E|)\)\\
Difference between chromatic number and greedy clique number\(^\P\) & \(O(|V|^{2})\)\\
\end{longtable}

\begin{flushleft}\footnotesize
\(^\dagger\)\;Computing all node degrees is \(O(|E|)\);
obtaining the median or standard deviation over the \(|V|\) values
requires sorting, for a total of \(O(|E|+|V|\log|V|)\).

\(^\ddagger\)\;Power iteration (or Lanczos) for the dominant eigenvector
needs \(k\) sparse matrix–vector products, each \(O(|E|)\);
overall \(O(k\,|E|)\).

\(^\S\)\;Using wedge–enumeration triangle counting; naive
algorithm cost is \(O(|V|^{3})\) which is still polynomial.

\(^\P\)\;The feature is implemented with
\texttt{greedy\_color(\dots)} followed by
\texttt{max\_clique(\dots)}.  
This feature is computed using Networkx library functions. The greedy “largest\_first’’ colouring scans every adjacency list once,
yielding \(O(|V|^{2})\) in the worst case; the clique heuristic
terminates quickly on sparse graphs and does not dominate the total
running time, so the combined procedure remains \(O(|V|^{2})\). It is a heuristic procedure and the exact chromatic number is NP-hard to compute.
\end{flushleft}
}

\subsection{Computational considerations}

The computational experiments were performed on a Linux Ubuntu 22.04.3 LTS 64 bit system, utilizing an AMD Ryzen 1920 processor with 24 threads operating at a clock speed of 4 GHZ and a TU106 GeForce RTX 2070 GPU equipped with 8GB of memory. Gurobi's version 12 was used as the solver and NetworkX version 2.8.4 was used to generate synthetic graphs and compute graph features. The CliSAT, MOMC and HGS algorithms required graphs to be converted into formats compatible with their respective implementations. While the pre-processing time needed for these conversions was not significant as compared to the algorithm runtimes, it was still excluded from runtime computation as it was a formatting and not a computational operation pertaining to solving the problem. For CliSAT, the total runtime was calculated as the sum of the values $ts$,$tr$ and $tp$ provided in the output files. However, it was observed that $ts$ was the dominant term in the vast majority of cases.

\subsection{Analysis of the compiled dataset}

\begin{figure}[h!]
    \centering
    \includegraphics[width=1.0\textwidth]{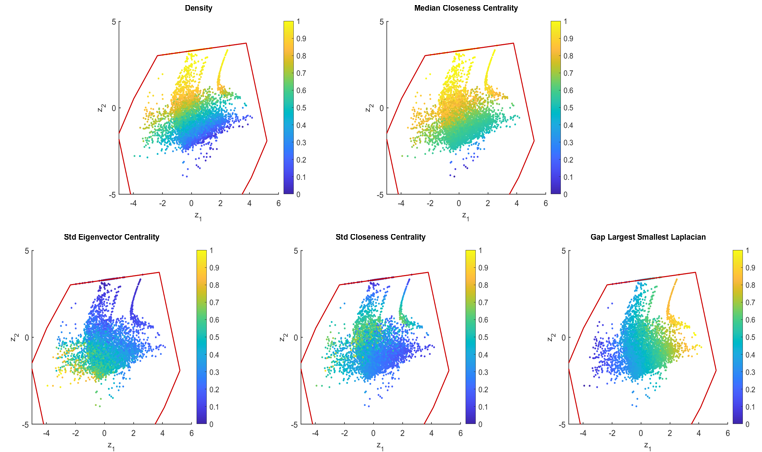}
    \caption{The features selected by the SIFTED routine of the ISA plotted on the $Z_{1}$-$Z_{2}$ plane for the existing dataset comprising of 6138 graph instances and the estimate of the boundaries of the instance space obtained by the CLOISTER routine (red boundary lines) }
    \label{fig:Z1Z2_projection_of_features_and_boundaries_train_dataset}
\end{figure}

The ISA methodology was applied to the compiled dataset of 6138 graph instances summarized in Table \ref{tab:summary_of_graphs_for_isa}. The value of the minimum required correlation coefficient between the features and the performance measure was selected as 0.8. The SIFTED auto-feature selection routine of ISA selected 5 of the 35 features based on the correlation coefficient value and the average silhouette value achieved during clustering. All these 5 features were generic graph-based features. The 2 problem-specific features did not have a correlation exceeding 0.8 with the defined performance measure. It, however, does not imply that such problem-specific features are not important as they may have a high correlation with other performance measures, may be selected as a part of a larger set of features if a lower correlation threshold were used, or may reveal other important insights about the structure of problems. The transformation relationship between the selected 5 graph features and the two-dimensional (2D) variables $Z_{1}$ and $Z_{2}$ is provided in Equation \ref{eq:feature_selection_and_transformation_existing_dataset}. Figure \ref{fig:Z1Z2_projection_of_features_and_boundaries_train_dataset} presents the contour plot of these features projected on the $Z_{1}$ -$Z_{2}$ plane. The plot of the feature \lq Gap Largest Smallest Laplacian\rq  reveals that the value of this  feature increases as $Z_{1}$ and $Z_{2}$ increase, aligning with the transformation matrix specified in Equation \ref{eq:feature_selection_and_transformation_existing_dataset} and similar visual checks can be performed for other features. The CLOISTER routine of ISA was used to estimate the boundaries of the instance space, depicted as red lines encasing the instances in these plots. It can be observed that while several regions of the instance space are well populated by the instances from the existing dataset, others, such as the region where $Z_{2} \leq -2.5$ are sparsely populated. While the focus of this study is ISA-based algorithm performance prediction on large instances and not on generating instances in sparse regions, this observation highlights the fact that even in datasets that are well-known in the research community, there can be some regions in the instance space where they may not be enough instances for understanding and comparing algorithm performance.    

\begin{equation} \label{eq:feature_selection_and_transformation_existing_dataset}
      \begin{bmatrix}
         Z_{1} \\
         Z_{2} \\ 
     \end{bmatrix}
     =
     \begin{bmatrix}
     
         -0.1653 & 0.6747 \\
          1.0134 & 0.1196 \\
          -0.1392 & 0.5377 \\
         -0.3048 & -0.4641 \\
          -0.2322 & -0.4569
     \end{bmatrix}^T
      \times
     \begin{bmatrix}
         \text{Density} \\
         \text{Gap Largest Smallest Laplacian} \\
         \text{Median Closeness Centrality} \\ 
         \text{Standard Deviation Closeness Centrality} \\
         \text{Standard Deviation Eigenvector Centrality}
         
     \end{bmatrix}
\end{equation}

\section{Results and discussion}
\label{sec:results}

\begin{figure}[h!]
    \centering
    \includegraphics[width=0.8\textwidth]{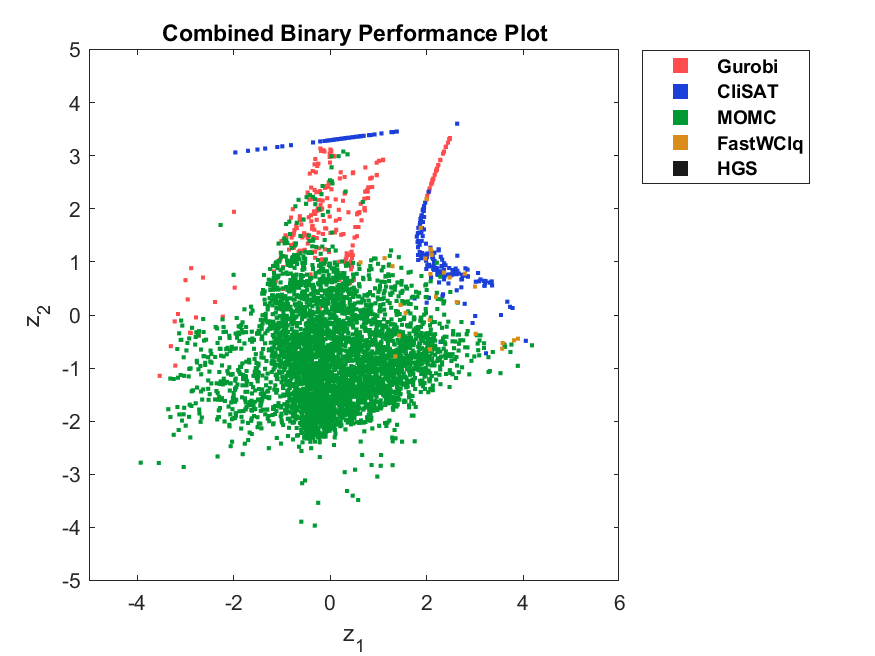}
    \caption{Instance space containing existing dataset containing instances from COLLAB, IMDB-BINARY and TWITTER graphs (6138 instances)}
    \label{fig:combined_algorithms_binary_performance_train}
\end{figure}

\begin{figure}[h!]
    \centering
    \includegraphics[width=1.0\textwidth]{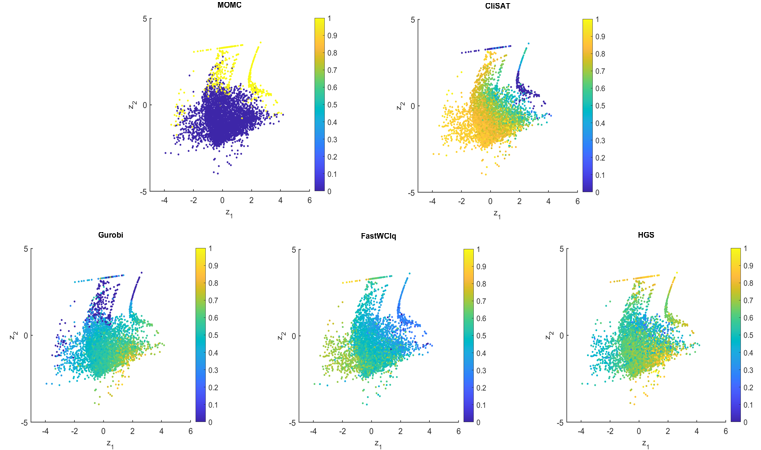}
    \caption{Performance of the five algorithms across all the instances of the existing dataset. The performance measure has been scaled from 0 to 1 and the scale applies across all the plots. A lower value of the performance measure indicates better performance.}
    \label{fig:algorithm_performance_measure_all_five_algorithms_train}
\end{figure}

Figure \ref{fig:combined_algorithms_binary_performance_train} shows the combined binary performance plot for all five algorithms over the instance space and Figure \ref{fig:algorithm_performance_measure_all_five_algorithms_train} shows the performance plot of the individual algorithms. Overall, MOMC emerged as the algorithm with the best performance with minimum $y_{min}$ scores in 4585 (74.7\%) of the 6138 instances. CliSAT, Gurobi and FastWClq yielded the best performance in 679 (11\%), 851 (13.8\%) and 23 (0.3\%) of the instances. HGS did not emerge as the best algorithm in any of the instances, but it did emerge as the second best in several cases.
On further analysis of the instance space, it was found that for each algorithm, there were regions where it performed better than others. MOMC had the best performance overall and it performed very well in the region ($Z_{2} \leq 1$) where graphs had densities and medians of closeness centralities in the low to medium range and standard deviations of their eigenvector centralities in the medium to high range. The tight color bounds of these sparse graphs and their easy-to-spot hubs in their hub-and-spoke structure are well-suited for MOMC's strategies of having few branches, identifying a strong early incumbent and performing massive pruning. Gurobi emerged as the best performing algorithm in the region of the instance space characterized by high densities and medians of closeness centralities and low to medium standard deviations of the closeness and eigenvalue centralities. In this region, graphs are densely connected, structurally uniform and globally small (short path lengths), properties that Gurobi’s presolve, symmetry handling and LP‑based bounding routines exploit to yield excellent performance. CliSAT also emerges as the best performing algorithm for instances that have the characteristics that enable Gurobi to excel with the distinct additional characteristic that the instances have low standard deviations of eigenvector centralities. Along with the benefits of encoding a sparse complement graph for the SAT problem and high symmetry, the absence of high‑influence hub nodes could likely yield tighter bounds and limit clause size, factors that may explain CliSAT's superior performance. While FastWClq and HGS attained best performances in too few instances to warrant a discussion on the possible factors, it is to be noted that several instances where FastWClq yielded best or at least good performance were present in the first quadrant ($Z_{1},Z_{2} \geq 0$) and for HGS, they were present in the first and particularly in the fourth ($Z_{1},Z_{2} \leq 0$) quadrants.       
               
\begin{figure}[H]
    \centering
    \includegraphics[width=0.8\textwidth]{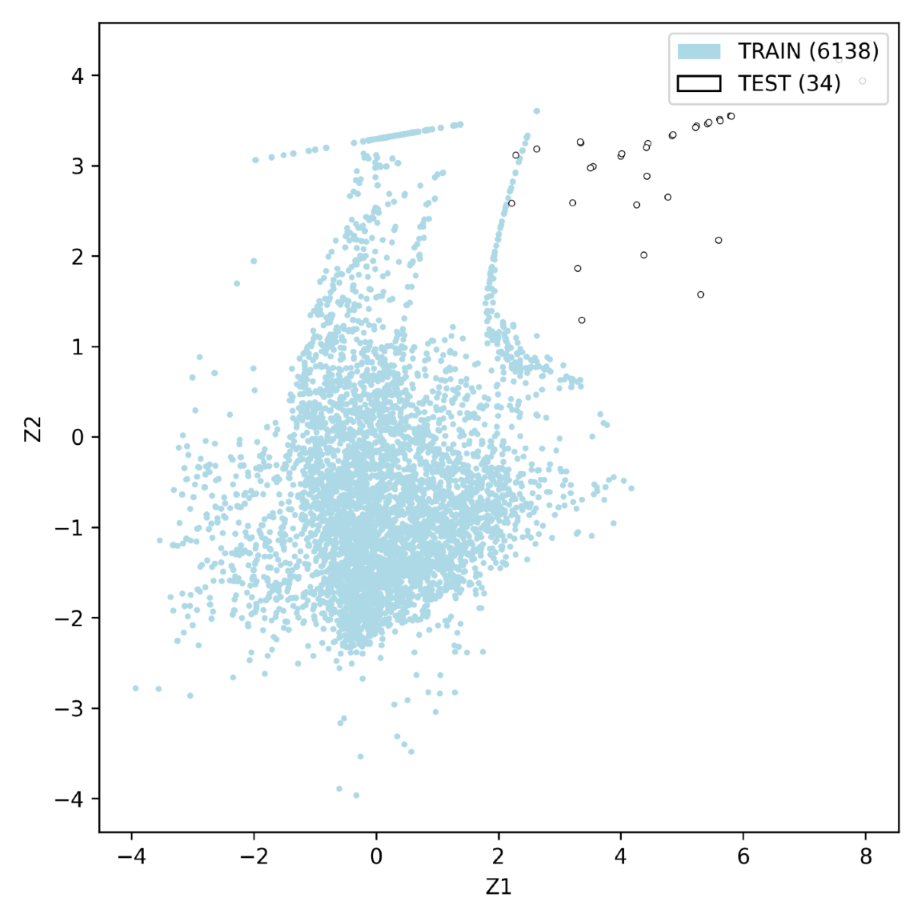}
    \caption{34 test instances from the BHOSLIB-DIMACS dataset projected on the Z1-Z2 plane along with the 6138 training instances compiled from TWITTER, IMDB-BINARY and COLLAB datasets.}
    \label{fig:test_instances_in_the_Z1Z2projection}
\end{figure}

The Support Vector Classifier (SVC) model trained by the PYTHIA module of the ISA was employed to forecast algorithm performance on a test set of substantially larger graph instances from the BHOSLIB and DIMACS datasets (\cite{bhoslib_dataset}). A total of 34 challenging instances were chosen from these datasets, including the two largest, frb100-40 (BHOSLIB) and MANN\_a81 (DIMACS). Table \ref{tab:test_graphs_stats_and_algo_performance} presents the node count, edge count, and density of these graph instances. Figure \ref{fig:test_instances_in_the_Z1Z2projection} illustrates the projection of these test instances alongside the training instances in 2D space, showing that most test instances are located in a distinct region compared to the training instances. Each of the five algorithms was executed for these instances, and the performance metric $y_{min}$ was calculated. A maximum runtime of 1800 seconds was set for all algorithms. The trained SVC model predicted the optimal algorithm for these instances, with results displayed in Table \ref{tab:test_graphs_stats_and_algo_performance}. The model was able to correctly predict the best performing algorithm for 30 out of these 34 challenging benchmark instances (see columns Actual best alg. vs Pred. best alg. in Table \ref{tab:test_graphs_stats_and_algo_performance}), which corresponds to an accuracy of 88\%. Also, the algorithm predicted by the model featured among the top 2 of the 5 algorithms for the test set 33 out of 34 times (see column Pred. best alg. in top-2 in Table \ref{tab:test_graphs_stats_and_algo_performance}), which corresponds to an accuracy of 97\%. The size of the cliques found by the actual best algorithm, predicted best algorithm, the largest clique found by any algorithm, the time taken by the actual best algorithm, and the time taken by the predicted best algorithm for each instance are also listed in Table \ref{tab:test_graphs_stats_and_algo_performance}. The model was able to correctly predict the best performing algorithms for instances that were in a part of the instance space different from the one it was trained on. Crucially, it was able to accurately predict that FastWClq and HGS were the best performing algorithms for 19 of these 34 instances even though they were the best performers for only 0.3\% and 0\% of the instances in the training data respectively. Essentially, the model was able to learn the mapping between graph features and algorithm performance for these algorithms and was able to predict their performance correctly on unseen instances even though it did not observe them as being best performers in the training dataset. It is to be noted that the instances in the training dataset were much smaller as compared to the instances in the test dataset. The median node counts for the training and the test datasets were 47 and 945 respectively. The largest graph in the training dataset (COLLAB\_4024) had 492 nodes and 37673 edges whereas the largest graph in the test dataset (BHOSLIB frb100-40) had 4000 nodes and 7425226 edges. The maximum time taken by any algorithm for any instance in the training set was 7.37 seconds and the total time taken by all the algorithms together for all the training instances was 2157.32 seconds whereas several algorithms touched the 1800 second time limit for the test instances and the cumulative time taken by all the algorithms together for all the test instances was 113305.22 seconds. Given such an extent of computational effort involved for solving large instances, the proposed methodology of training the ISA-based performance prediction model on a dataset of smaller instances to accurately predict best performing algorithms before actually solving problem instances offers significant savings of computational effort for researchers as well as practitioners for selecting algorithms for solving large scale combinatorial optimization problems.                           

\newgeometry{
  left=1.1cm,
  right=1.1cm,
  top=1.1cm,
  bottom=1.1cm
}

\begin{sidewaystable}[p]
\scriptsize                                   
\setlength{\tabcolsep}{3pt}                   
\renewcommand{\arraystretch}{0.9}             
\centering
\caption{Benchmark graphs, summary statistics and instance‑space‑analysis
algorithm performance prediction for the \textsc{BHOSLIB}–\textsc{DIMACS} graph instances}
\label{tab:test_graphs_stats_and_algo_performance}
\begin{tabular}{@{}llrrr
  C{1.65cm} C{1.65cm} C{1.65cm}               
  C{1.65cm} C{1.65cm} C{1.65cm}               
  C{1.65cm} C{1.65cm}@{} }                    
  
\toprule
\textbf{Set} & \textbf{Graph} & \textbf{\#V} & \textbf{\#E} & \textbf{Dens.} &
\makecell{\textbf{Actual}\\\textbf{best}\\\textbf{alg.}} &
\makecell{\textbf{Pred.}\\\textbf{best}\\\textbf{alg.}} &
\makecell{\textbf{Pred.}\\\textbf{best alg.}\\\textbf{in top‑2}} &
\makecell{\textbf{Clique}\\\textbf{actual}\\\textbf{(best alg.)}} &
\makecell{\textbf{Clique}\\\textbf{pred.}\\\textbf{(best alg.)}} &
\makecell{\textbf{Largest}\\\textbf{clique}\\\textbf{(any alg.)}} &
\makecell{\textbf{Time(s)}\\\textbf{actual}\\\textbf{best alg.}} &
\makecell{\textbf{Time(s)}\\\textbf{pred.}\\\textbf{best alg.}} \\
\midrule
BHOSLIB & frb30-15-1           &  450 &   83\,198 & 0.8235 & MOMC & MOMC & yes & 30 & 30 & 30 & 0.21 & 0.21 \\
BHOSLIB & frb30-15-5           &  450 &   83\,231 & 0.8239 & MOMC & MOMC & yes & 30 & 30 & 30 & 0.24 & 0.24 \\
BHOSLIB & frb35-17-1           &  595 &  148\,859 & 0.8424 & MOMC & MOMC & yes & 35 & 35 & 35 & 0.86 & 0.86 \\
BHOSLIB & frb35-17-2           &  595 &  148\,868 & 0.8424 & FastWClq & FastWClq & yes & 33 & 33 & 35 & 1.03 & 1.03 \\
BHOSLIB & frb40-19-4           &  760 &  246\,815 & 0.8557 & MOMC & HGS & yes & 40 & 28 & 40 & 4.46 & 21.93 \\
BHOSLIB & frb40-19-1           &  760 &  247\,106 & 0.8568 & MOMC & MOMC & yes & 40 & 40 & 40 & 1.61 & 1.61 \\
BHOSLIB & frb45-21-1           &  945 &  386\,854 & 0.8673 & HGS & HGS & yes & 32 & 32 & 45 & 42.35 & 42.35 \\
BHOSLIB & frb45-21-5           &  945 &  387\,461 & 0.8687 & HGS & HGS & yes & 32 & 32 & 45 & 42.26 & 42.26 \\
BHOSLIB & frb50-23-3           & 1150 &  579\,607 & 0.8773 & FastWClq & FastWClq & yes & 46 & 46 & 49 & 10.22 & 10.22 \\
BHOSLIB & frb50-23-4           & 1150 &  580\,417 & 0.8785 & MOMC & FastWClq & yes & 50 & 46 & 50 & 11.80 & 59.09 \\
BHOSLIB & frb53-24-1           & 1272 &  714\,129 & 0.8834 & FastWClq & HGS & yes & 48 & 39 & 50 & 91.46 & 129.58 \\
BHOSLIB & frb53-24-5           & 1272 &  714\,130 & 0.8834 & FastWClq & HGS & yes & 48 & 39 & 52 & 81.69 & 129.45 \\
BHOSLIB & frb56-25-5           & 1400 &  869\,699 & 0.8881 & FastWClq & HGS & yes & 52 & 39 & 56 & 210.79 & 192.99 \\
BHOSLIB & frb56-25-2           & 1400 &  869\,899 & 0.8883 & HGS & HGS & yes & 41 & 41 & 54 & 192.90 & 192.90 \\
BHOSLIB & frb59-26-2           & 1534 &1\,049\,648 & 0.8927 & FastWClq & FastWClq & yes & 53 & 53 & 57 & 36.75 & 36.75 \\
BHOSLIB & frb59-26-5           & 1534 &1\,049\,829 & 0.8929 & FastWClq & FastWClq & yes & 54 & 54 & 57 & 87.61 & 87.61  \\
BHOSLIB & frb100-40            & 4000 &7\,425\,226 & 0.9284 & FastWClq & HGS & no & 87 & 71 & 93 & 704.86 & 1800 \\
DIMACS  & brock200\_1          &  200 &   14\,834 & 0.7454 & FastWClq & FastWClq & yes & 21 & 21 & 21 & 0.03 & 0.03 \\
DIMACS  & sanr200\_0.9         &  200 &   17\,863 & 0.8976 & FastWClq & FastWClq & yes & 42 & 42 & 42 & 0.1 & 0.1 \\
DIMACS  & C250.9               &  250 &   27\,984 & 0.8991 & FastWClq & FastWClq & yes & 44 & 44 & 44 & 0.14 & 0.14 \\
DIMACS  & sanr400\_0.7         &  400 &   55\,869 & 0.7001 & FastWClq & FastWClq & yes & 21 & 21 & 21 & 0.21 & 0.21 \\
DIMACS  & dsjc500.5            &  500 &   62\,624 & 0.5020 & FastWClq & FastWClq & yes & 13 & 13 & 13 & 0.1 & 0.1 \\
DIMACS  & p\_hat500-2          &  500 &   62\,946 & 0.5046 & CliSAT & CliSAT & yes & 36 & 36 & 36 & 0.08 & 0.08 \\
DIMACS  & gen400\_p0.9\_65     &  400 &   71\,820 & 0.9000 & CliSAT & CliSAT & yes & 65 & 65 & 65 & 0.17 & 0.17 \\
DIMACS  & gen400\_p0.9\_75     &  400 &   71\,820 & 0.9000 & CliSAT & CliSAT & yes & 75 & 75 & 75 & 0.09 & 0.09 \\
DIMACS  & brock800\_1          &  800 &  207\,505 & 0.6493 & FastWClq & HGS & yes & 21 & 14 & 23 & 8.96 & 18.73 \\
DIMACS  & keller5              &  776 &  225\,990 & 0.7515 & FastWClq & FastWClq & yes & 27 & 27 & 27 & 1.21 & 1.21 \\
DIMACS  & dsjc1000.5           & 1000 &  249\,826 & 0.5002 & FastWClq & FastWClq & yes & 15 & 15 & 15 & 2.78 & 2.78 \\
DIMACS  & p\_hat1000-3         & 1000 &  371\,746 & 0.7442 & HGS & HGS & yes & 26 & 26 & 68 & 42.07 & 42.07 \\
DIMACS  & hamming10-2          & 1024 &  518\,656 & 0.9902 & FastWClq & FastWClq & yes & 512 & 512 & 512 & 0.02 & 0.02 \\
DIMACS  & MANN\_a45            & 1035 &  533\,115 & 0.9963 & CliSAT & CliSAT & yes & 345 & 345 & 345 & 4.09 & 4.09 \\
DIMACS  & p\_hat1500-2         & 1500 &  568\,960 & 0.5061 & FastWClq & FastWClq & yes & 65 & 65 & 65 & 29.90 & 29.90 \\
DIMACS  & C2000.5              & 2000 &  999\,836 & 0.5002 & FastWClq & FastWClq & yes & 16 & 16 & 16 & 7.03 & 7.03 \\
DIMACS  & MANN\_a81            & 3321 &5\,506\,380& 0.9988 & FastWClq & FastWClq & yes & 1082 & 1082 & 1100 & 282.93 & 282.93 \\
\bottomrule
\end{tabular}
\end{sidewaystable}

\restoregeometry        
\section{Conclusion and further research }
\label{sec:conclusion}

This study systematically applies the ISA methodology to evaluate the performance of SOTA exact, heuristic and GNN-based algorithms for the MCP. A data set comprising 6138 instances was compiled from the existing benchmarks of the COLLAB, IMDB-BINARY, and TWITTER data sets. For each instance, 33 generic graph-based features and 2 problem-specific features were computed, and the largest computed cliques sizes were obtained using five algorithms, with their respective runtimes recorded. A composite performance measure that incorporates both the size of the computed clique and the runtime was used to evaluate algorithm performance. The features and performance measures were processed through the ISA pipeline to obtain the projections of the instances onto the 2D \mbox{$Z_1$-$Z_2$} space and to visualize the comparative performance of the algorithms in the instance space. It was found that MOMC outperformed other algorithms in 74.7\% of instances, while Gurobi and CliSAT exhibited superior performance in specific regions of the instance space and were the best performing algorithms in 13.8\% and 11\% of the instances respectively. A test dataset of 34 challenging instances from the DIMACS and BHOSLIB datasets was compiled and the SVC model trained on the 6138 instances was applied on them to determine the best performing algorithm for each of them. These instances were substantially larger and lay on a separate part of the instance space as compared to the instances of the training dataset. The top-1 and top-2 prediction accuracy of the model for the best performing algorithm was 88\% and 97\% respectively. The model was able to learn the mappings between graph features and algorithm performance and was able to predict FastWClq and HGS as the best performing algorithms for 19 of these 34 test instances even though they were the best performers for only 0.3\% and 0\% of the instances in the training set respectively. Given the extent of computational effort required to solve large combinatorial optimization problems, the proposed methodology offers insights into problem hardness as well as significant time savings for researchers and practitioners.\\

This work opens up multiple avenues for further research. First, future studies could further investigate the specific properties of graph instances that lead to superior performance by HGS, CliSAT, and Gurobi in distinct regions of the instance space. This study described the features of the instances in these different regions qualitatively and linked them to the solution strategies of these algorithms that may be enabling them to achieve superior performance. Quantitative studies using feature values can be performed to confirm whether these hypotheses are valid. Second, the inclusion of quantum computing algorithms, such as the Quantum Approximate Optimization Algorithm (QAOA) based on the Quadratic Unconstrained Binary Optimization (QUBO) framework, could provide an interesting comparison with exact, heuristic and GNN-based methods. Third, since the graph features employed in this study are generic and problem agnostic, they can be used to conduct ISA to evaluate algorithm performance for other graph-based COPs such as minimum vertex cover, graph coloring, minimum dominating set, etc. that do not involve distances or have directionality between nodes. Fourth, with modifications to include distance-based graph-level features such as an adjacency matrix representing distances, the proposed approach can also be potentially extended for graph-based COPs that have distances between nodes such as the TSP, VRP, LRP, etc. Fourth, while a fixed high value of the correlation coefficient (0.8) was used in the SIFTED routine to select the features that were most correlated with the defined performance measure, it would be interesting to vary this coefficient and analyze its effect on the number of features selected by the routine and the performance of the prediction model. Finally, the performance measure used in this study, while effective, is one of many possible formulations. For example, a threshold for the maximal clique size (say set at 80\% of the largest clique size computed) could be added to the performance measure to address situations where an algorithm may run fast but compute significantly smaller clique sizes than other algorithms. Additionally, alternative evaluation functions, such as polynomial formulations with weighted components for the maximal clique size and runtime, could be explored. The authors anticipate investigating some of these research directions in future work and hope that the findings presented here will inspire further research on comparative algorithm performance for graph-based COPs.\\

\newpage

\section*{Declarations}

\begin{itemize}

\item \textbf{CRediT authorship contribution statement}: \textbf{Bharat S. Sharman}: Conceptualization, Methodology, Software, Validation, Formal analysis, Investigation, Data curation, Writing – original draft, Writing – review \& editing, Visualization. \textbf{Elkafi Hassini}: Conceptualization, Methodology, Project administration, Resources, Supervision, Writing – review \& editing.
\item \textbf{Compliance with Ethical Standards}: 
\begin{enumerate}
    \item \textbf{Funding}: The authors did not receive support from any organization for the submitted work.
    \item \textbf{Conflict of Interest}:  Bharat Sharman declares that he has no conflict of interest. Elkafi Hassini declares that has no conflict of interest.
    \item \textbf{Interests}: The authors have no relevant financial or non-financial interests to disclose.
    \item \textbf{Ethical Approval}: This article does not contain any studies with human participants or animals performed by any of the authors.
\end{enumerate}

\item \textbf{Data Availability}
The datasets analyzed during the current study are available in the following repositories: 
\begin{enumerate}
    \item TWITTER: (http://snap.stanford.edu/data)
    \item COLLAB and IMDB-BINARY: (https://chrsmrrs.github.io/datasets/docs/datasets/) 
    \item BHOSLIB and DIMACS: (https://networkrepository.com/)

\end{enumerate} 
\item \textbf{Code availability}: The code developed for this study will be made publicly available through GitHub upon publication. 
\item \textbf{Declaration of generative AI and AI-assisted technologies in the writing process}: During the preparation of this work the author(s) used Writefull for scientific writing to paraphrase their initial draft to improve its scientific language. After using this tool, the author(s) reviewed and edited the content as needed and take(s) full responsibility for the content of the published article.  
\end{itemize}

\bmhead{Acknowledgements}

The authors acknowledge Gurobi Optimization LLC, for providing free access to the academic license. The authors also acknowledge the authors of the CliSAT, MOMC, HGS \& FastWClq algorithms and the ISA framework who either kindly shared their source code with the authors or made them openly available, which was essential for this research. \\
 


\end{document}